\documentclass{article}

\usepackage[margin=1in]{geometry} 
\usepackage[hidelinks]{hyperref}
\usepackage{xcolor}
\hypersetup{
    colorlinks,
    linkcolor={red!50!black},
    citecolor={blue!50!black},
    urlcolor={blue!80!black}
}

\usepackage{graphicx}
\usepackage{floatrow}
\usepackage{adjustbox}
\usepackage[labelformat=simple]{caption, subfig}
\usepackage{longtable}
\usepackage{booktabs}
\usepackage{array}
\usepackage{enumitem}
\usepackage{appendix}
\usepackage{titling}
\usepackage[noblocks]{authblk} 
\usepackage[symbol*,multiple]{footmisc} 
\setfnsymbol{chicago}

\usepackage[mathlines]{lineno}

\usepackage{mathtools}
\usepackage{amssymb}        
\usepackage{mathabx}
\usepackage[amsmath, amsthm, thmmarks]{ntheorem}
\usepackage[capitalize, noabbrev]{cleveref}
\crefname{equation}{eq.}{eqs.}
\crefname{appendix}{Supplementary Information}{Supplementary Information}

\usepackage[citestyle=authoryear,style=authoryear,uniquename=false,doi=false,isbn=false,url=false,eprint=false]{biblatex}
\renewcommand{\cite}{\textcite}
\newcommand{\citep}{\parencite}

\DeclareMathOperator{\Var}{Var}

\newcommand*\patchAmsMathEnvironmentForLineno[1]{%
  \expandafter\let\csname old#1\expandafter\endcsname\csname #1\endcsname
  \expandafter\let\csname oldend#1\expandafter\endcsname\csname end#1\endcsname
  \renewenvironment{#1}%
     {\linenomath\csname old#1\endcsname}%
     {\csname oldend#1\endcsname\endlinenomath}}%
\newcommand*\patchBothAmsMathEnvironmentsForLineno[1]{%
  \patchAmsMathEnvironmentForLineno{#1}%
  \patchAmsMathEnvironmentForLineno{#1*}}%
\AtBeginDocument{%
\patchBothAmsMathEnvironmentsForLineno{equation}%
\patchBothAmsMathEnvironmentsForLineno{align}%
\patchBothAmsMathEnvironmentsForLineno{flalign}%
\patchBothAmsMathEnvironmentsForLineno{alignat}%
\patchBothAmsMathEnvironmentsForLineno{gather}%
\patchBothAmsMathEnvironmentsForLineno{multline}%
}

\title{Cultural transmission of move choice in chess}

\author[1,$\dagger$]{Egor Lappo}
\author[1]{Noah A.\ Rosenberg}
\author[1]{Marcus W.\ Feldman}

\affil[1]{Department of Biology, Stanford University, Stanford, CA 94305 USA}
\affil[$\dagger$]{Email: \href{mailto:elappo@stanford.edu}{\texttt{elappo@stanford.edu}}}

\date{\today}

\begin{document}

\maketitle

\noindent {\bf Abstract.} 
The study of cultural evolution benefits from detailed analysis of cultural transmission in specific human domains. Chess provides a platform for understanding the transmission of knowledge due to its active community of players, precise behaviors, and long-term records of high-quality data. In this paper, we perform an analysis of chess in the context of cultural evolution, describing multiple cultural factors that affect move choice. We then build a population-level statistical model of move choice in chess, based on the Dirichlet-multinomial likelihood, to analyze cultural transmission over decades of recorded games played by leading players. For moves made in specific positions, we evaluate the relative effects of frequency-dependent bias, success bias, and prestige bias on the dynamics of move frequencies. We observe that negative frequency-dependent bias plays a role in the dynamics of certain moves, and that other moves are compatible with transmission under prestige bias or success bias. These apparent biases may reflect recent changes, namely the introduction of computer chess engines and online tournament broadcasts. Our analysis of chess provides insights into broader questions concerning how social learning biases affect cultural evolution.

\medskip


\section{Introduction}\label{sec:intro}

Chess has existed in its current form for hundreds of years; it is beloved as an established sport, a hobby, and also as a source of inspiration for scientists across disciplines. Since the 1950s, playing chess well has served as a goal in the development of artificial intelligence, as a task that a ``thinking agent'' would be able to accomplish \citep{shannon1950chess}. This goal was realized in the victory of a chess algorithm over a top human player (Deep Blue vs. Garry Kasparov in 1997). In physics and signal processing, researchers study time series in databases of chess games to extract information regarding long-term correlations, dynamics of position evaluation, invention of new openings, and other game features \citep[see e.g.][]{schaigorodsky2016study, blasius2009zipf, ribeiro2013move, perotti2013innovation}. Statisticians have been interested in chess as a case study in the development of human performance measurement \citep{regan2011understanding, di2009skill} and modeling of human choice \citep{regan2014human, regan2014human}. 

As a \emph{cultural} dataset, a compendium of chess games has great potential to help cultural evolution researchers understand patterns of cultural transmission and social learning. A large body of well-annotated chess games is available online, and, unlike linguistic or textual data, for example, these data contain a precise record of players' behavior. As chess positions and moves are discrete, they can be recorded with complete information. Yet the space of potential game sequences is extremely large, so that there can be great variation in move choices. In addition, the large amount of canonical literature on chess allows for thorough qualitative interpretation of patterns in move choice.

Focusing on the game of Go, a game that also features discrete moves and complete information, \cite{beheim2014strategic} analyzed the choice of the first move by Go players in a dataset of $\sim$31,000 games. They concluded that the choice of the first move is driven by a mix of social and individual factors, and the strength of these influences depends on the player's age. 
Many issues concerning cultural transmission in board games remain to be studied. For example, what are the mechanisms behind social learning: are players choosing to use ``successful'' moves or, instead, moves played by successful players? What defines success of a move? Answering these questions contributes to understanding both general processes of the spread of innovations and mechanisms that govern dynamics of the evolution of cultural traits.

In this paper, we perform a quantitative study of chess in the context of cultural evolution using a database of 3.45 million chess games from 1971 to 2019. In \cref{sec:game info}, we introduce chess vocabulary and several aspects of the game important for our analysis. In \cref{sec:culture}, we describe cultural factors involved in the game and position them within the context of existing literature on cultural transmission. 
\Cref{sec:data} describes the dataset used in this study. 
In \cref{sec:move choice}, we motivate and define a statistical model for frequencies of opening strategies in the dataset. Unlike individual-based analysis of a binary choice of the first move in Go by \cite{beheim2014strategic}, our model incorporates counts for all possible moves in a position, taking a population-level approach. In \cref{sec:model analysis}, we discuss the fit of the model to data for three positions at different depths in the game tree.

\section{The game of chess}\label{sec:game info}

In this section, we briefly review chess vocabulary, assuming readers have some basic knowledge of the rules of the game \citep[for a concise summary, see][]{capablanca2002primer}.

First, a game of chess consists of two players taking turns moving one of their pieces on the board, starting with the player who is assigned the white pieces. We will call these discrete actions \emph{plys}: the first ply is a move by the white player, the second ply is a move by the black player, and so on. The average length of a chess game at a professional level is around $80$ plys (see \cref{sec:data} below). We will use the word ``ply'' when describing specific positions, but otherwise we will use the words ``move,'' ``strategy,'' and ``response'' interchangeably with ``ply.''

Moves are typically recorded using \emph{algebraic notation} \citep[p.\ 389]{oxford_companion}, in which each ply is represented by a letter for a piece --- \textbf{K} for king, \textbf{Q} for queen, \textbf{R} for rook, \textbf{B} for bishop, \textbf{N} for knight, no letter for a pawn --- followed by the coordinates of the square on which the piece ends. The coordinates on the board are recorded using letters from \textbf{a} to \textbf{h} from left to right for the \emph{ranks} (the $x$-axis coordinates), and numbers from 1 to 8 for the \emph{files} (the $y$-axis coordinates). For example, the first few moves of the game could be recorded as \textbf{1.\ e4 e5 2.\ Nf3 Nc6 3.\ Bc4 Nf6\ldots} Other special symbols are used for captures (\textbf{x}), checks (\textbf{+}), and castling (\textbf{O-O} or \textbf{O-O-O} for king- and queen-side castling, respectively).

\begin{figure}[bht]\centering
\sidesubfloat[]{\includegraphics[width = 0.25\textwidth]{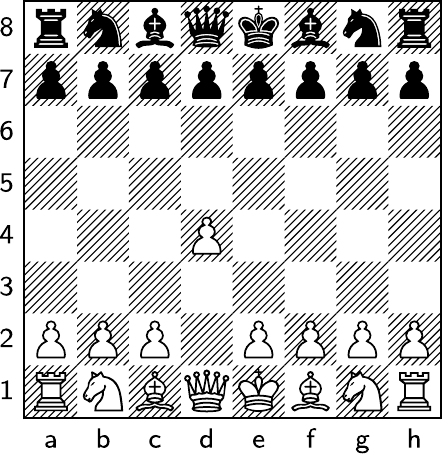}\label{subfig:position example a}}\hfill
\sidesubfloat[]{\includegraphics[width = 0.25\textwidth]{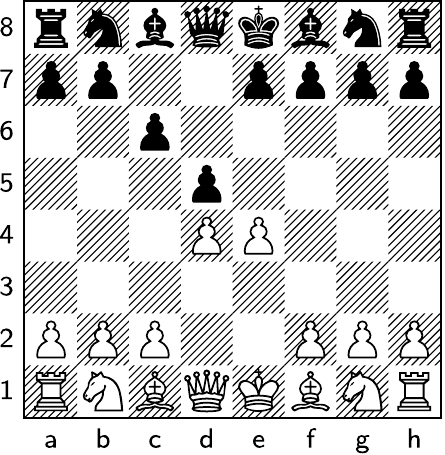}\label{subfig:position example b}}\hfill
\sidesubfloat[]{\includegraphics[width = 0.25\textwidth]{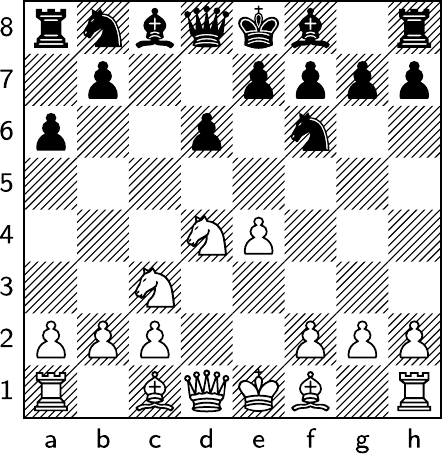}\label{subfig:position example c}}    
\caption{Example chess opening positions. (A) Queen's Pawn opening, \textbf{1.\ d4}. (B) Caro-Kann opening, \textbf{1.\ e4 c6 2.\ d4 d5}. \ (C) Najdorf Sicilian opening, \textbf{1.\ e4 c5 2.\ Nf3 d6 3.\ d4 cxd4 4.\ Nxd4 Nf6 5.\ Nc3 a6}.}
\label{fig:position example}     
\end{figure}

The initial stage of the game is called the \emph{opening}. In the opening, players try to achieve a favorable arrangement of the pieces that gives them the most freedom for further actions while keeping their kings safe. Openings are highly standardized, with many having names, e.g. the Sicilian Defense, or the London Opening. Because the number of possible positions is not that large at the beginning of the game, openings are extensively analyzed by players and then memorized for use in tournaments. Example chess positions in the opening are presented in \cref{fig:position example}.

The collective body of knowledge about how to play chess from various positions is called \emph{chess theory}. For the opening, theory consists of extensive analyses of  many positions by human players as well as by computers. One of the manifestations of chess theory is the existence of fixed sequences of moves called ``lines,'' from which deviations are rare. A \emph{mainline} is a sequence of moves that has proven to be the most challenging for both opponents, such that neither of them is able to claim an advantage. A \emph{sideline} is a sequence of moves that deviates from the established optimal sequence. 

Each professional chess player has a numerical rating, usually assigned by the national or international federation. FIDE (The International Chess Federation) uses the \emph{Elo rating system} \citep{elo}. The rating is relative, meaning that it is calculated based on a player's past performance, and is intended to represent a measure of the player's ability. 
The typical rating of a strong intermediate player is $\sim$1500, and a rating of $2500$ is required to qualify for a Grandmaster (GM) title. Most elite tournaments involve ratings above $2700$.

\section{Culture and chess}\label{sec:culture}
Chess is a cultural practice that is actively shaped by the people who participate in it. Individual players enter the practice, altering their performance and behaviors depending on the games they and others have played. 
Many cultural processes are involved in players' decision-making. To analyze these processes, we will concentrate on decisions made in the opening stage, because the relatively small number of positions allows players to reason about concrete moves and lines in their analyses and preparation. The factors affecting move choice that we discuss below are well-known to the chess community \citep{euwe_style, desjarlais2011counterplay,gobet2018psychology, de2002chess}. Our goal here is to place them in the language of cultural evolution.

\begin{enumerate}[label=(\alph*)]
    \item \textbf{Objective strength.} One factor in move choice is the objective strength of the move, which reflects the potential for victory from resulting positions. An evaluation of a move's strength can be made by human analysis or with a chess computer. 
    Many early moves have been extensively analyzed, so the best choice in those positions is well-known to most professional players. 

    \item \textbf{Social context of the move.} 
    Players are aware of how often a given move has been played in the past. This frequency evaluation can even be automated using websites such as \href{https://www.openingtree.com}{\texttt{OpeningTree.com}}. Developed theory often exists for more frequent moves, which can be the default choice for many players. Conversely, rare moves or \emph{novelties} (previously unseen moves) can create problems for opponents who most likely have not prepared a response.

    It is important to observe that the frequency with which a move is played is not directly proportional to the objective strength discussed in (a); there are moves that are objectively weak, but only conditional on the opponent finding a \emph{single} good response. If this response is not played by the opponent, then the weak move may give an advantage. In some conditions, e.g.\ an unprepared opponent or lack of time, such a ``weak'' move can be highly advantageous. There have also been cases in which a historically frequent move was later ``refuted'' by deep computer analysis.

    Beyond the move frequency, information on the success of strategies in leading to a win can play a role in move choice. In many positions, actually applying information about objective move strength is a complex problem. It is not enough to make a single strong move: a player must then \emph{prove} an advantage by continuing to play further strong moves and executing plans that would lead to victory. The success rate of a move is an indicator of how hard it is to gain a long-term advantage leading to checkmate after choosing it. 
    
    The influence of elite players may also be important in move choice. Top players participate in invitational tournaments followed by the wider community. Players, presented with a choice of approximately similar moves, may choose the one that was played by a ``superstar'' player. This phenomenon is exemplified by strategies named after famous players, such as ``Alekhine's Defense'' \citep[p. 159]{mco15} or ``Najdorf Sicilian'' \citep[p. 246]{mco15}. Leading players can create trends; for example, the Berlin defense was popularized after grandmaster Vladimir Kramnik employed it to win the World Championship in 2000 \citep[p. 43]{mco15}. 

    \item \textbf{Metastrategy.} Beyond trends in move choice, the ``metastrategy'' of chess is also evolving. Conceptions of what a game of chess ``should'' look like have been changing through the years, and so has the repertoire of openings used by professional players \citep[p.\ 359]{oxford_companion}. 
    In the 18th century, the swashbuckling Romantic style of chess emphasized winning with ``style'': declining gambits, or offers of an opponent's piece, could be viewed as ungentlemanly, and Queen's Pawn openings were rarely played \citep[Ch.\ 5]{shenk2011immortal}.
    However, by the World Championship of 1927, trends in chess had shifted to long-term positional play \citep[see][Ch.\ 8]{shenk2011immortal}. Queen's Pawn openings were the cutting edge of chess theory, and almost all games at that tournament began with the Queen's Gambit Declined \citep{1927wc}.
    Following World War I, hypermodern chess emphasized control of the board's center from a distance, and its influence is evident in top-level games of the mid-20th century \citep[Ch.\ 10]{shenk2011immortal}. Hypermodern players refused to commit their pawns forward, preferring a position where pieces are placed on safe squares from which they could target the opponent's weaknesses.
    Recently, a style of chess mimicking computer play has emerged, in which players memorize long computer-supported opening lines and play risky pawn advances.

    Chess is as much a social phenomenon as it is individual. Some players exhibit personal preferences for certain game features, such as early attacks or long and complicated endgames, and some aspects of play are determined by a player's upbringing. For example, the Soviet school of chess formed around a certain energetic, daring, and yet ``level-headed'' style \citep{kotov1961soviet}.
    
    \item \textbf{Psychological aspects.} Finally, psychological aspects and circumstances of the game contribute to move choice \citep{gobet2004psychology}. There are lines that are known to lead to a quick draw, and a player might elect to follow one of them, depending on the relevance of the outcome at a particular stage of the tournament. Openings may also be chosen to take opponents out of their comfort zone: in a game against a much weaker opponent, a dynamic and ``pushy'' line might give a player an advantage. Similarly, a master of attacking play might make mistakes when forced into a long positional game.
\end{enumerate}

The complexities of move choice suggest that chess could serve as a model example for the quantitative study of culture. Players' knowledge is continually altered by their own preparation, the games they play, and other players' actions. In this sense, chess knowledge is ``transmitted'' over time, in part by players observing and imitating their own past actions and those of other players, or \emph{transmission by random copying} \citep{bentley2007regular}. The large historical database of chess games provides an opportunity to study deviations from random copying dynamics known as \emph{transmission biases} or \emph{social learning strategies} \citep{boyd_richerson, kendal2018social, laland2004social, henrich2003evolution}. In our analysis of the transmission of chess knowledge, we will investigate \emph{success bias} (players paying attention to win rates of different strategies), \emph{prestige bias} (players imitating the world's best grandmasters), and \emph{frequency-dependent bias} (e.g.\ players choosing rare or unknown strategies). 

\section{Data}\label{sec:data}

The dataset that serves as the foundation for this project is \emph{Caissabase} -- a compendium of $\sim$5.6 million chess games, available for download at \href{http://caissabase.co.uk}{\texttt{caissabase.co.uk}}. Games in the dataset involve players with Elo rating $2000$ or above, and correspond to master-level play, allowing us to focus on the dynamics of high-level chess without the influence of players who are just learning the game. 

In filtering the dataset, we have excluded games with errors that did not correspond to a valid sequence of moves as determined by a chess notation parser. We also filtered the dataset to keep only the games that record the result of the game, players' names, and their Elo ratings, and we selected only the games played from $1971$ to $2019$. This filtering produced a table of 3,448,853 games.

\begin{figure}[tbh]
    \centering
    \includegraphics[width=0.4\textwidth]{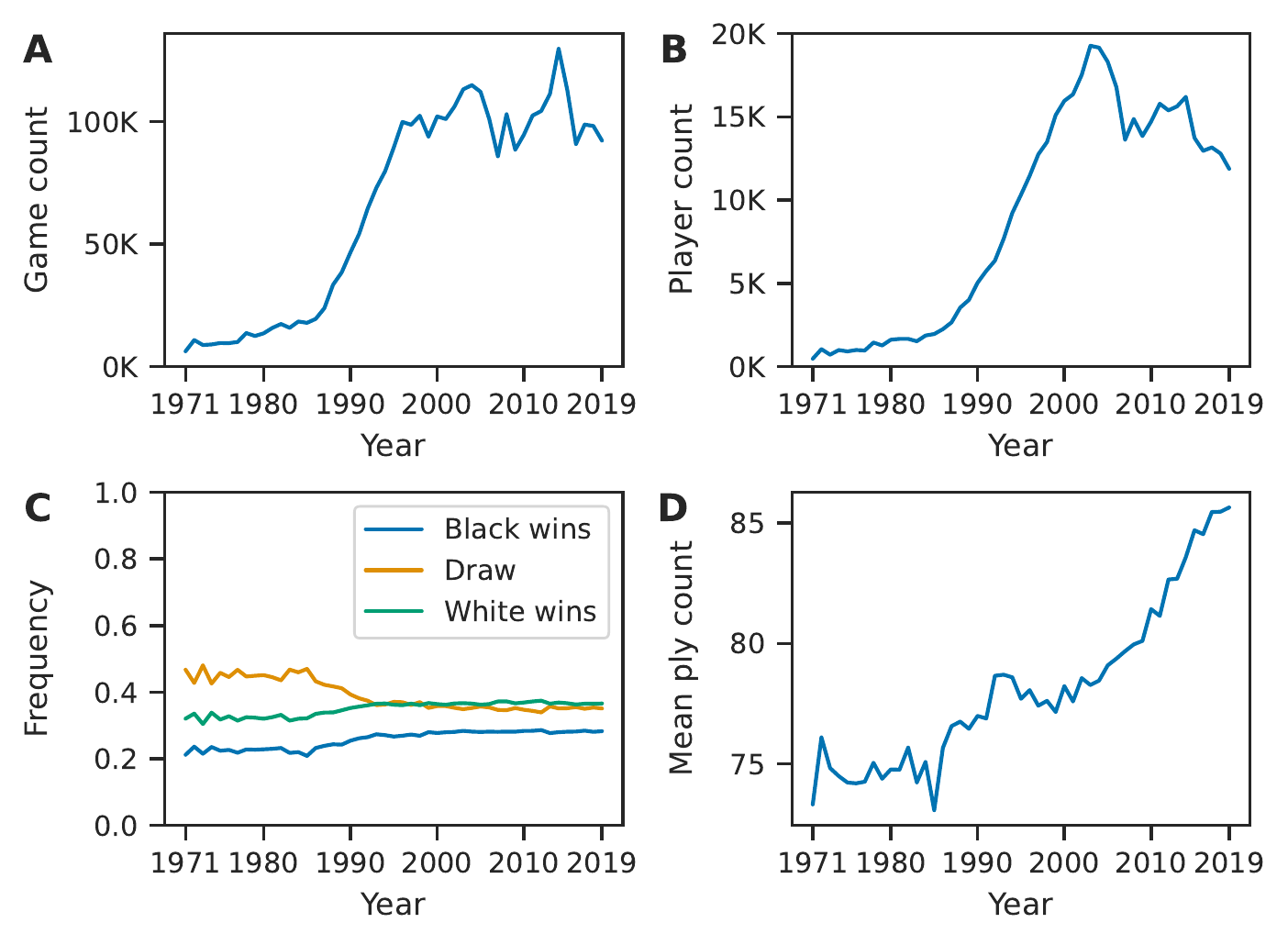}
    \caption{Features of the dataset. (A) Number of games per year. (B) Number of unique players per year. (C) Outcome proportions in each year. (D) Average game length per year, measured in the number of plys (half-moves).}
    \label{fig:dataset plots}
\end{figure}

In \Cref{fig:dataset plots}, we highlight the main aspects of the dataset. \Cref{fig:dataset plots}A shows that the number of games per year has been growing steadily since the $1970$s, stabilizing at approximately 100,000.
In total, there are 77,956 chess players in the dataset, with the number of players per year increasing in recent decades (\cref{fig:dataset plots}B). 

It is widely accepted in the chess community that white has a slight advantage, as the side that starts the game. This view is reflected in \cref{fig:dataset plots}C, which plots the fractions of outcomes of games in each year. Finally, \cref{fig:dataset plots}D shows the average length of games over time; games have become longer since the mid-1980s, which could mean that players are getting better at the game and no longer lose early.
To explore the dynamics in the dataset further, we examine the frequencies of individual moves.

\section{Modeling move choice}\label{sec:move choice}

\subsection{Move frequencies}\label{subsec:streamplots}

\begin{figure}[tbh]
    \centering
    \includegraphics[width=0.8\textwidth]{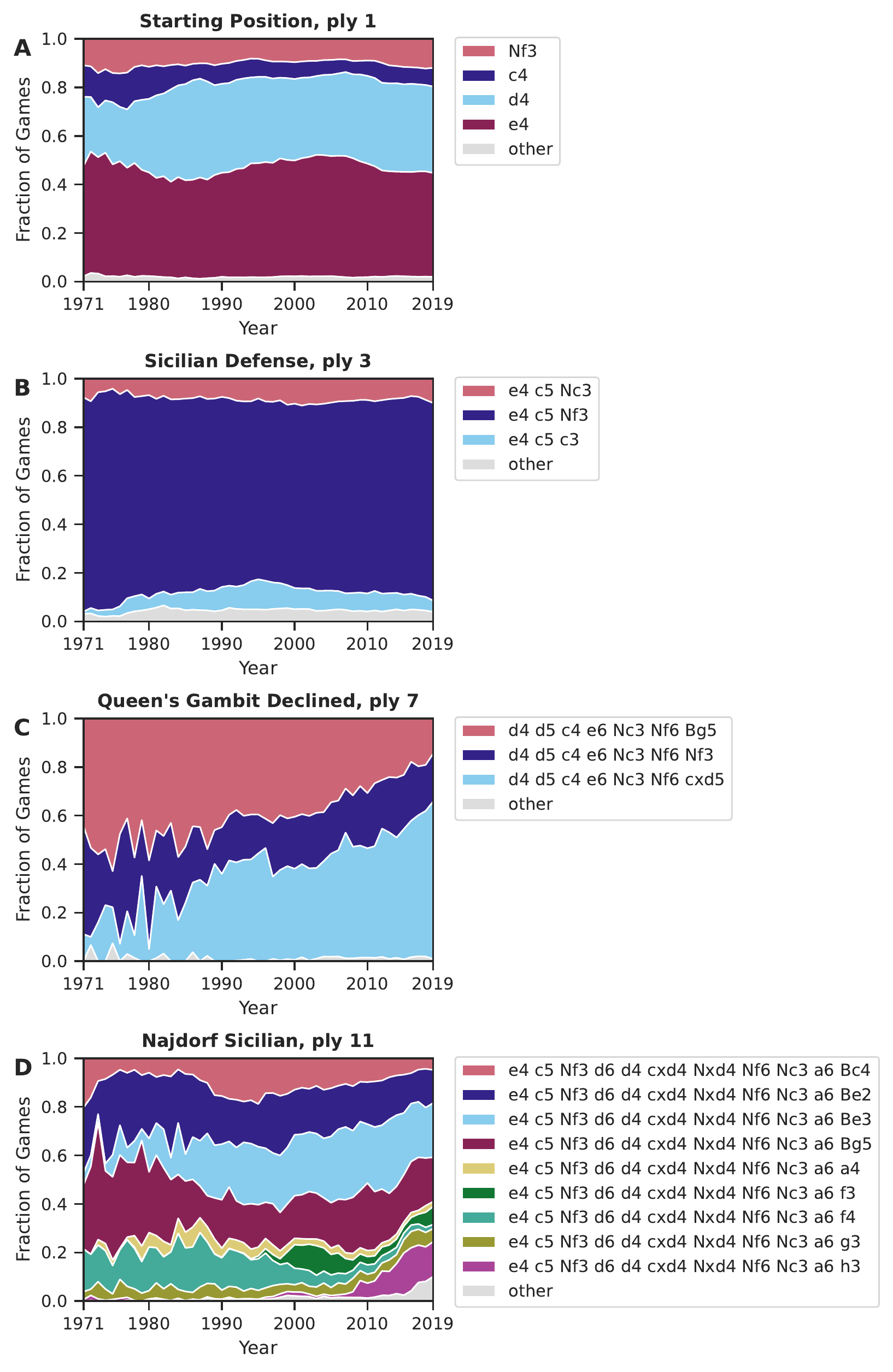}
    \caption{Move frequencies over time. For each panel, the legend presents the whole sequence of moves from the start of the game, with odd moves played by white, and even moves by black. The ``other'' category contains all rare moves that individually have average frequency less than 2\%, with the average taken over all years. (A) Starting Position. (B) Sicilian Defense. (C) Queen's Gambit Declined. (C) Najdorf Sicilian. See the main text for a discussion of each position.}
    \label{fig:streamplots}
\end{figure}

Here, we discuss the dynamics of move frequencies over time for several game positions. Given a position on the board, the player whose turn it is has a choice of which move to play. In positions where their king is in check, players would only have few choices, since they are forced to get out of check. In some other cases, several equally attractive moves could be available, and any of the factors in \cref{sec:culture} has the potential to affect the choice.
Depending on the position, the move frequency trajectories look drastically different, as shown in \cref{fig:streamplots}.

\textbf{Starting Position, ply 1.} \Cref{fig:streamplots}A shows the fractions of games in which different starting moves were played in each year from $1971$ to $2019$. 
The frequencies of the moves are mostly constant over time, suggesting that the starting move is a well-understood and well-developed idea. 

\textbf{Sicilian Defense, ply 3.}  
\Cref{fig:streamplots}B shows move frequencies in response to \textbf{1.\ e4 c5} --- the Sicilian Defense. In this position, there is a mainline move --- \textbf{Nf3} --- which an overwhelming majority of players prefer to play, while other moves are rarely played. Move distributions in which one specific move dominates are common, possibly because some sequences of moves are perceived as a single coherent unit. 

\textbf{Queen's Gambit Declined, ply 7.} \Cref{fig:streamplots}C presents an example of a gradual change, which might have happened either due to a change in the metastrategy of play or because of the gradual development of chess theory.

\textbf{Najdorf Sicilian, ply 11.} A game starting with a Sicilian Defense can follow a sequence known as the Najdorf Sicilian. This sequence consists of 10 plys, and the moves at ply 11 that have been played in the resulting position are presented in \cref{fig:streamplots}D. Qualitatively, the picture is dramatically different from the early positions considered above. Among the responses to the Najdorf Sicilian, some moves are consistently popular choices (\textbf{Be2}, \textbf{Be3}, \textbf{Bg5}), some became ``obsolete'' in recent years (\textbf{f4}), and some rapidly gained popularity (\textbf{h3}).

The qualitative picture of move frequency changes can be summarized as follows. On one hand, very early opening moves do not show large fluctuations in frequencies, most likely because a significant change in frequency necessitates some kind of ``innovation,'' which is impossible to produce at such an early stage. On the other hand, moves beyond the standardized opening frequencies (after the 16th-20th ply) involve positions that do not repeat often enough for humans to memorize and analyze during preparation. This property makes quantitative analysis of specific late-game moves nearly impossible. Somewhere between these two extremes are positions at which chess theory is actively developed and tested. Positions such as the Najdorf Sicilian occur early enough in the game to be reached often, but are advanced enough to provide many continuation possibilities that are approximately equal in terms of objective strength. In such positions, all factors, including engine analysis, move frequency, social context, stylistic trends, and personal preferences could play a role in move choice. 

\subsection{Population-level modeling of move choice}\label{subsec:model}

We develop a statistical model that can help to understand the data described above. 
A complete model of move choice would involve parameters associated with the whole population, with subgroups of players (e.g.\ top 50 players), or with each individual. Such a model would be very complex, so our model is restricted to population-level features of dynamics; we analyze frequency-dependent, success, and prestige biases. Features concerning match-level dynamics, personal development, and preferences of individual players are outside of the scope of our analysis, and are present in the form of residual variance, not explained by our population-level treatment.

\subsubsection{Unbiased model}

First, we consider a null model that generates the simplest dynamics, reflecting unbiased transmission of move choice preferences from one year to the next. Conceptually, the model assumes that each year, players ``sample'' a move randomly from games that were played in the last year. More precisely, fix an arbitrary chess position and suppose that in each year $t$, exactly $N_t$ games having this position were played. The data for the model are the counts of $k$ different response moves, denoted by $\boldsymbol{x}_t = (x^1_{t},\ldots,x^k_{t})$. We do not attempt to model appearance of novel strategies, so we will assume that all counts are positive, $x^i_t > 0$. The vector of response strategy counts in the next year, $\boldsymbol{x}_{t+1}$, is multinomially distributed,

\begin{equation}
    \boldsymbol x_{t+1} \sim \text{Multinomial}(N_{t+1}, \boldsymbol{\theta}_t).
\end{equation}

\noindent The probability vector $\boldsymbol{\theta}_t$ has the Dirichlet distribution with counts in the current year, $\boldsymbol{x}_t$, as Dirichlet allocation parameters,

\begin{equation}
    \boldsymbol{\theta}_t \sim \text{Dirichlet}(\boldsymbol x_t).
\end{equation}

The multinomial likelihood depends on a positive integer parameter $n$ and a vector of probabilities $\boldsymbol\theta$ that sum to one,

\begin{equation}
    f_M(\boldsymbol y; n, \boldsymbol{\theta}) = \frac{n!}{y_1!\cdots y_k!}\theta_1^{y_1}\cdots \theta_k^{y_k}; \sum_{i=1}^k y_i = n.
\end{equation}
The Dirichlet likelihood depends on a vector of positive real numbers $\boldsymbol\alpha$:

\begin{equation}
    f_D(\boldsymbol{\theta}; \boldsymbol{\alpha}) = \frac{\Gamma\left(\sum_{i=1}^k \alpha_i\right)}{\prod_{i=1}^k \Gamma(\alpha_i)} \prod_{i=1}^k \theta_i^{\alpha_i-1}.
\end{equation}
These two likelihoods can be combined into the compound Dirichlet-multinomial likelihood by integrating over $\boldsymbol{\theta}$ \citep[pp.\ 80-83]{johnson1997discrete},

\begin{equation}\label{eq:dm likelihood}
    f_{DM}(\boldsymbol y; n, \boldsymbol\alpha) = \frac{n!\,\Gamma\left(\sum_{i=1}^k \alpha_i\right)}{\Gamma\left(n + \sum_{i=1}^k \alpha_i\right)} \prod_{i=1}^k \frac{\Gamma(y_i + \alpha_i)}{y_i!\,\Gamma(\alpha_i)},
\end{equation}
which will be the likelihood for the model. In other words, under our unbiased model, the counts $\boldsymbol x_{t+1}$ of moves in year $t+1$ are distributed with probability density function 

\begin{equation}
p(\boldsymbol x_{t+1} \mid N_{t+1}, \boldsymbol x_t) = f_{DM}(\boldsymbol x_{t+1}; N_{t+1}, \boldsymbol x_{t}),
\end{equation}
so that the counts in the previous year $\boldsymbol x_t$ take the roles of the Dirichlet parameters $\boldsymbol \alpha$.
As a shorthand, we write

\begin{equation}\label{eq:null model shorthand}
    \boldsymbol x_{t+1} \sim \text{Dirichlet-multinomial}(N_{t+1}, \boldsymbol{x}_t).
\end{equation}

For a vector $\boldsymbol y$ having a Dirichlet-multinomial distribution with parameters $n$ and $\boldsymbol \alpha$, the expectation is

\begin{equation}
    \mathbb E\left[\boldsymbol y \right] = \frac{n}{\sum_{j=1}^k \alpha_j}\boldsymbol\alpha.
\end{equation}
For our model, this formula yields

\begin{equation}\label{eq:null model expect}
    \mathbb E\left[\boldsymbol{x}_{t+1}\right] = \frac{N_{t+1}}{\sum_{j=1}^k x^j_{t}}\boldsymbol x_t  = \frac{N_{t+1}}{N_t} \boldsymbol x_t,    
\end{equation}
meaning that no changes are expected to happen in this unbiased model, except possibly for the change in the number of games played. The strategies are ``transmitted'' from one year to the next proportionally to their current frequencies in the population. 

The null model is analogous to a neutral many-allele Wright--Fisher model in population genetics \citep{ewens}. The multinomial distribution arises as a representation of a biological process in Wright--Fisher models, where individuals in the next generation ``choose'' a parent from the previous generation. In our model of move choice, such sampling is a metaphor that does not correspond exactly to an observed physical process. As we discuss below, working with counts directly via the Dirichlet distribution allows us to account for a potentially higher variance in the strategy counts relative to the multinomial distribution \citep{corsini2022dealing}. Use of the Dirichlet-multinomial likelihood is a common way of dealing with overdispersion in count data in many fields, including ecology \citep{harrison2020dirichlet} and microbiome studies \citep{wadsworth2017integrative,osborne2022latent}.

It should be noted that chess players pay attention to games further back in the past than just the last year. Our null model is still a reasonable representation of the process for several reasons. First, there is a high degree of autocorrelation in the move count data \citep{schaigorodsky2016study}, meaning that it is likely that the most recent data point is representative of counts in the last several years. Second, players tend to look only at \emph{select} famous games of the past, whereas the more recent games can be more easily perceived in their totality.

\subsubsection{Fitness and frequency-dependence}\label{subsubsec:model:fitness}

A strategy transmitted at a rate greater than expected from the null model can be said to have higher \emph{cultural fitness} \citep{cavalli_feldman}. Conversely, a strategy having a lower transmission rate than expected has lower cultural fitness. Selection on strategies is carried out by players when they decide which move to play based on any of the factors discussed in \cref{sec:culture}. We can account for cultural fitness by associating a \emph{fitness coefficient} $f_i$ to  each strategy $i$. For now, assume that fitness values are constant, $0<f_i<\infty$. The distribution of moves in the next year can then be described as

\begin{equation}
     \boldsymbol x_{t+1} \sim \text{Dirichlet-multinomial}(N_{t+1}, f_1 x^1_t,\ldots,f_kx^k_t),
\end{equation}
with the expression for expected counts in the next year becoming

\begin{equation}\label{eq:fitness expectation}
    \mathbb E\left[x^i_{t+1}\right] = N_{t+1} \frac{f_ix^i_t}{\sum_{j=1}^k f_jx^j_t}.
\end{equation}
As the coefficients $f_i$ are constrained only in that they must be positive, this way of encoding the parameters is useful for inference purposes, especially in the  Bayesian framework we employ below. It is straightforward to find reasonable prior distributions on $(0,\infty)$, and absence of ``sum to one'' constraints makes it easy for an MCMC sampler to efficiently explore the posterior distribution \citep[Ch.~12]{bda}.

However, interpretation of the model is more convenient with a different parameterization: instead of considering values of $f_i$, we let

\begin{equation}\label{eq:fbar}
  \bar f_t = \frac{1}{N_t}\sum_{j=1}^k f_j\,x_t^j
\end{equation}
be the \emph{mean fitness} at time $t$, and define
\begin{equation}\label{eq:fprime}
f_i' = f_i/\bar f_t
\end{equation}
to be normalized fitness coefficients, such that $\sum_{i=1}^k f_i' = 1$. Rewriting \cref{eq:fitness expectation} as

\begin{equation}
    \mathbb E\left[x^i_{t+1}\right] = \frac{N_{t+1}}{N_t}\frac{f_i}{\bar f_t} x^i_t = \frac{N_{t+1}}{N_t}f'_i x^i_t,
\end{equation}
we see that $f_i' = 1$ implies no expected change in the frequency of strategy $i$ from time $t$ to $t+1$. Therefore, this choice of parameterization allows us to view $f_i'$ as growth rates, with $f_i' = 1$ corresponding to no selective advantage, i.e.\ the neutral case. The value of $\bar f_t$, in turn, adjusts the variance of the counts in the next year.

To summarize, in our Dirichlet-multinomial model, the $f_i$'s measure two phenomena at once; their \emph{relative} values represent selection, while the \emph{mean} value of the $f_i$'s measures overdispersion with respect to the multinomial model. Mathematically, the expectation of a $\text{Dirichlet}(\boldsymbol\alpha)$-distributed random variable is invariant with respect to multiplying $\boldsymbol\alpha$ by a positive constant, but its variance is determined by the magnitudes of the parameters. Although the $f_i$ are convenient to use in inference, we will interpret the results in terms of a parameterization that involves $f_i'$ and $\bar f_t$ (\cref{eq:fbar,eq:fprime}).

We now allow $f_i$ to depend on the frequency of the strategy, such that 

\begin{equation}
     \boldsymbol x_{t+1} \sim \text{Dirichlet-multinomial}(N_{t+1}, f_1(x^1_t/N_t) \,x^1_t,\ldots,f_k(x^k_t/N_t)\,x^k_t).
\end{equation}
In this way, we are able to incorporate \emph{frequency-dependent selection} phenomena, which have previously been shown to be present in models of cultural data \citep[e.g.\ ][]{plotkin2022}. Hence, we will refer to $f_i$ as \emph{frequency-dependent fitness functions}. The expression for the mean fitness now becomes

\begin{equation}\label{eq:fbar t}
    \bar f_t = \sum_{j=1}^k f_j(p^j_t)\, p^j_t,
\end{equation}
where $p^j_t = x^j_t/N_t$, and $k$ is the number of distinct moves played from a position.

We choose a piecewise-constant form for the functions $f_i$, as this form introduces minimal assumptions about their shape while keeping the number of parameters low. That is, for $i=1,\ldots,k$, we have

\begin{equation}\label{eq:fi def}
    f_i(x) = \begin{cases}
        c^i_1 & \text{ if }x \in [0,b^i_{1}), \\
        c^i_j & \text{ if }x \in [b^i_{j-1},b^i_{j}), \\
        c^i_\ell & \text{ if }x \in [b^i_{\ell-1},1], \\
    \end{cases}
\end{equation}
where $c^i_{j}$ are values of $f_i$ and $b^i_{j}$ are breakpoints that determine the boundaries of constant segments. 
For $\ell$ segments, $\ell-1$ breakpoints $b^i_j\in (0,1)$ must be specified.
We choose quartiles of move frequencies as the values for $b^i_{j}$, so that each function $f_i$ has three breakpoints and $\ell=4$ constant segments. This choice does not uniformly cover the domain of $f_i$, but allows for the same amount of data to be used in estimating each segment. 

\subsubsection{Full model}\label{subsec:model def}

We complete our model by accounting for additional features that could affect move choice dynamics.
In the final model, the vector of strategy counts in the year $t+1$ again has the Dirichlet-multinomial distribution with parameters $N_{t+1}$ and $\boldsymbol \alpha$:

\begin{equation}\label{eq:lk1}
    \boldsymbol x_{t+1} \sim \text{Dirichlet-multinomial}(N_{t+1},\boldsymbol\alpha).
\end{equation}
However, vector $\boldsymbol \alpha$ is now defined as 

\begin{equation}\label{eq:model alpha}
    \alpha_i = \exp(\boldsymbol \beta_i \cdot \boldsymbol y^i_t)\, f_i(x^i_t/N_t)\, x^i_{t}.
\end{equation}
Here, $x^i_t$ is the count of games with the $i$th strategy in year $t$, $f_i$ is a piecewise constant function of the strategy frequency described in \cref{subsubsec:model:fitness} above, and $\boldsymbol{\beta}_i$ is a vector of constant coefficients. 

Additional features beyond just the move count or frequency are denoted $\boldsymbol y_t^i$ in \cref{eq:model alpha}. There are three of these features:
\begin{enumerate}
    \item The average outcome of the strategy in the whole population for games in year $t$, with a win for the side making the move encoded as $1$, a win for the opposing side encoded as $-1$, and a draw encoded as $0$. We denote the corresponding coefficient by $\beta_{\text{win},i}$.
    \item The average outcome of the strategy among the top 50 players in the dataset in year $t$, encoded in the same way as the population win rate. The list of top 50 players was compiled separately for each year using the average Elo rating of the players in that year. We denote the corresponding coefficient by $\beta_{\text{top50-win},i}$.
    \item The frequency of the strategy among the top 50 players in year $t$. We denote the corresponding coefficient by $\beta_{\text{top50-freq},i}$.
\end{enumerate}
These features represent biases different from frequency dependence that could also contribute to cultural fitness of moves; if the average outcome significantly affects move choice, success bias is present in transmission, as represented by coefficients $\beta_{\text{win},i}$ and $\beta_{\text{top50-win},i}$. Similarly, prestige bias could be important for transmission if players imitate the top 50 players  as represented by coefficients $\beta_{\text{top50-win},i}$ and $\beta_{\text{top50-freq},i}$.

The extra features are included in the model as an exponential factor $\exp(\boldsymbol \beta_i \cdot \boldsymbol y^i_t)$. This choice of factor has two purposes: first, it ensures that the variables $\alpha_i$ stay positive for all parameter values and data points; second, it represents \emph{multiplicative} effects of several types of transmission biases, a common approach both in theoretical models of cultural evolution \citep[see e.g.\ ][]{denton2020cultural, fpm_conformity} and in analyses of experimental data \citep{barrett2017pay,deffner2020dynamic,canteloup2021processing}.

\subsubsection{Inference}\label{subsec:model inference}

In total, the parameter vector $\boldsymbol \theta = (c^i_j, \boldsymbol \beta_i)$ has length $7k$, where $k$ is the number of different moves played in a given position. For each move, there are three coefficients $\beta_{\text{win}, i}$, $\beta_{\text{top50-win}, i}$, $\beta_{\text{top50-freq}, i}$, as well as four values $c^i_{1},c^i_2, c^i_3,c^i_{4}$ characterizing the function $f_i$ in \cref{eq:fi def}. 

We choose to fit the model in a Bayesian framework using Markov Chain Monte Carlo sampling, as this choice makes implementation of the model straightforward and allows us to obtain both point estimates and uncertainty quantification from the same analysis. To conduct Bayesian inference, we need to specify a prior distribution for $\boldsymbol \theta$. Following \cite{bda}, we specify non-informative priors for each parameter. Each constant segment $c^i_j$ of each function $f_i$ was assigned an $\text{Exp}(1)$ prior, such that $f_i$ is always non-negative, and the prior mean of $f_i$ is equal to one, corresponding to neutrality. We assigned each parameter $\boldsymbol\beta_i$ a normal $\mathcal N(0,1)$ prior and standardized the corresponding features $\boldsymbol{y}_t^i$ to have zero mean and unit variance. Given these priors and the model likelihood (defined in \cref{eq:lk1,eq:model alpha}), samples were generated from the posterior distribution using the Hamiltonian Markov Chain Monte-Carlo sampler provided by the Stan software package \citep{gelman2015stan,stan}. For this procedure, we only consider the data from 1980 to 2019, since earlier years have significantly less data available.

Many moves were played only a few times in the whole dataset. To prevent extremely rare moves from inflating the number of parameters, we have combined moves that individually have average frequency less than 2\% into a single category called ``other.'' In addition, it is commonly accepted by professional players that rare moves serve the same purpose: to take the opponent ``out of theory'' into positions where neither player had spent significant time preparing, leading to more chaotic and tense games. 

There are also years in which some move counts are equal to zero, and in this case, our assumption that move counts are nonzero is violated. To remedy this situation, in computational inference we replace the parameter $\boldsymbol\alpha$ from \cref{eq:model alpha} by $\boldsymbol\alpha +1$, such that for all strategies,

\begin{equation}\label{eq:model alpha mod}
    \alpha_i = 1 + \exp(\boldsymbol \beta_i \cdot \boldsymbol y^i_t) f_i(x^i_t/N_t) x^i_{t}.
\end{equation}
This approach is commonly used to deal with the potential for zero counts of rare categories in models involving multinomial likelihoods. For example, it is used in Dirichlet-multinomial modeling of ecological data \citep{harrison2020dirichlet} and in multinomial ``assignment tests'' of individuals to populations in genetics \citep{paetkau1995microsatellite,rosenberg2005algorithms}.
For moves with non-zero counts, this correction biases expectations from $x^i_t/N_t$ to $(x^i_t + 1)/(N_t + K)$, where $K$ is the number of strategies. The bias is negligible when move counts are in the hundreds or above.

\section{Modeling results}\label{sec:model analysis}

We discuss model fits for three positions at three different depths in the game tree: the Queen's Pawn opening at ply 2 (\textbf{1.\ d4}), the Caro-Kann opening at ply 5 (\textbf{1.\ e4 c6 2.\ d4 d5}), and the Najdorf Sicilian at ply 11 (\textbf{1.\ e4 c5 2.\ Nf3 d6 3.\ d4 cxd4 4.\ Nxd4 Nf6 5.\ Nc3 a6}). The parameters of the Stan HMC sampler and convergence diagnostics for each position are reported in \cref{supp:stan info}. In total, there are $N=1,083,146$ games with the Queen's Pawn opening, $N=80,890$ games with the Caro-Kann opening, and $N=82,557$ games with the Najdorf Sicilian opening. Input data such as raw strategy counts and win rates in each year appear in \cref{supp:input features}.

\begin{figure}[bht]\centering
\sidesubfloat{\includegraphics[width = 0.32\textwidth]{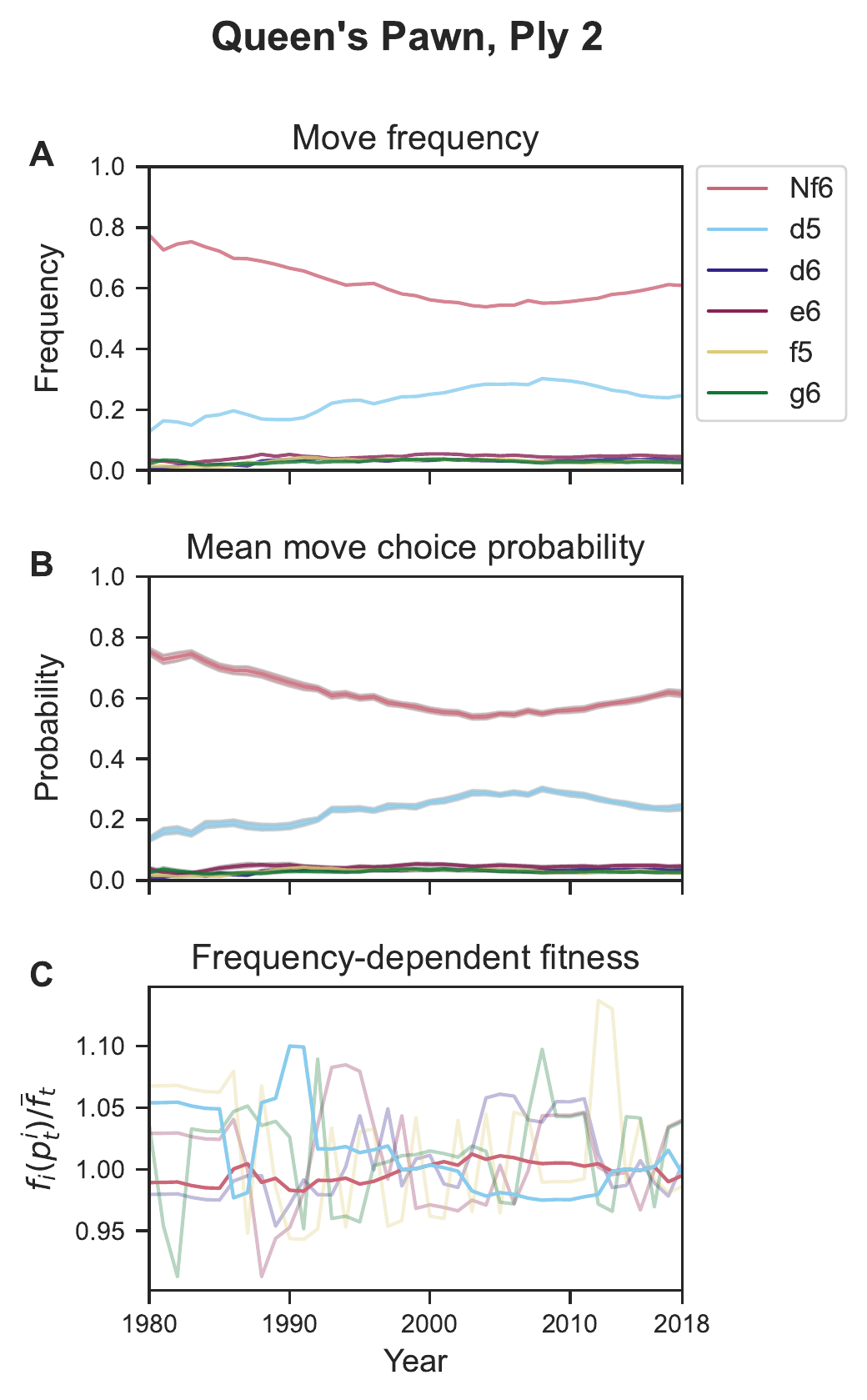}\label{subfig:model a}}\hfill
\sidesubfloat{\includegraphics[width = 0.32\textwidth]{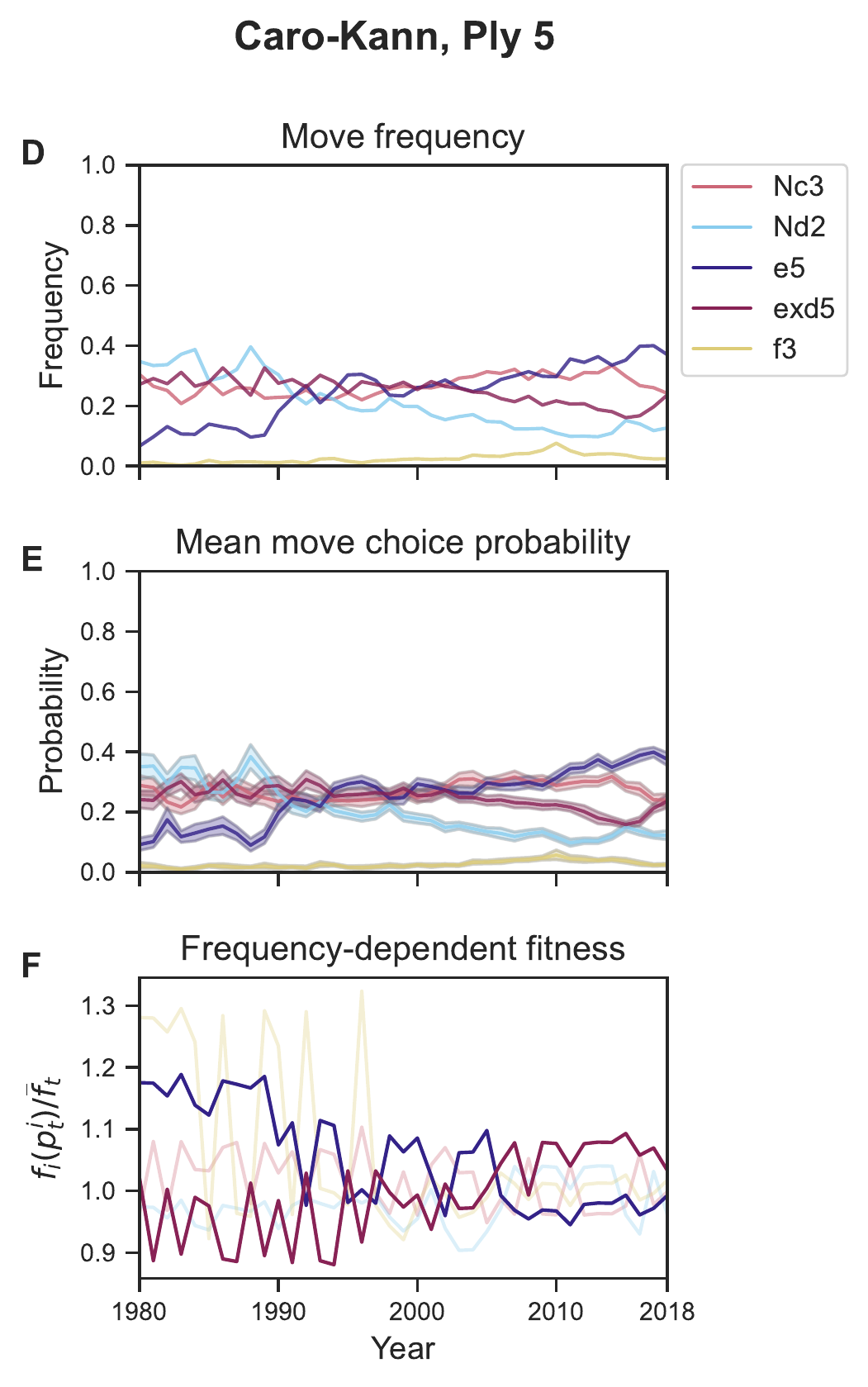}\label{subfig:model b}}\hfill
\sidesubfloat{\includegraphics[width = 0.32\textwidth]{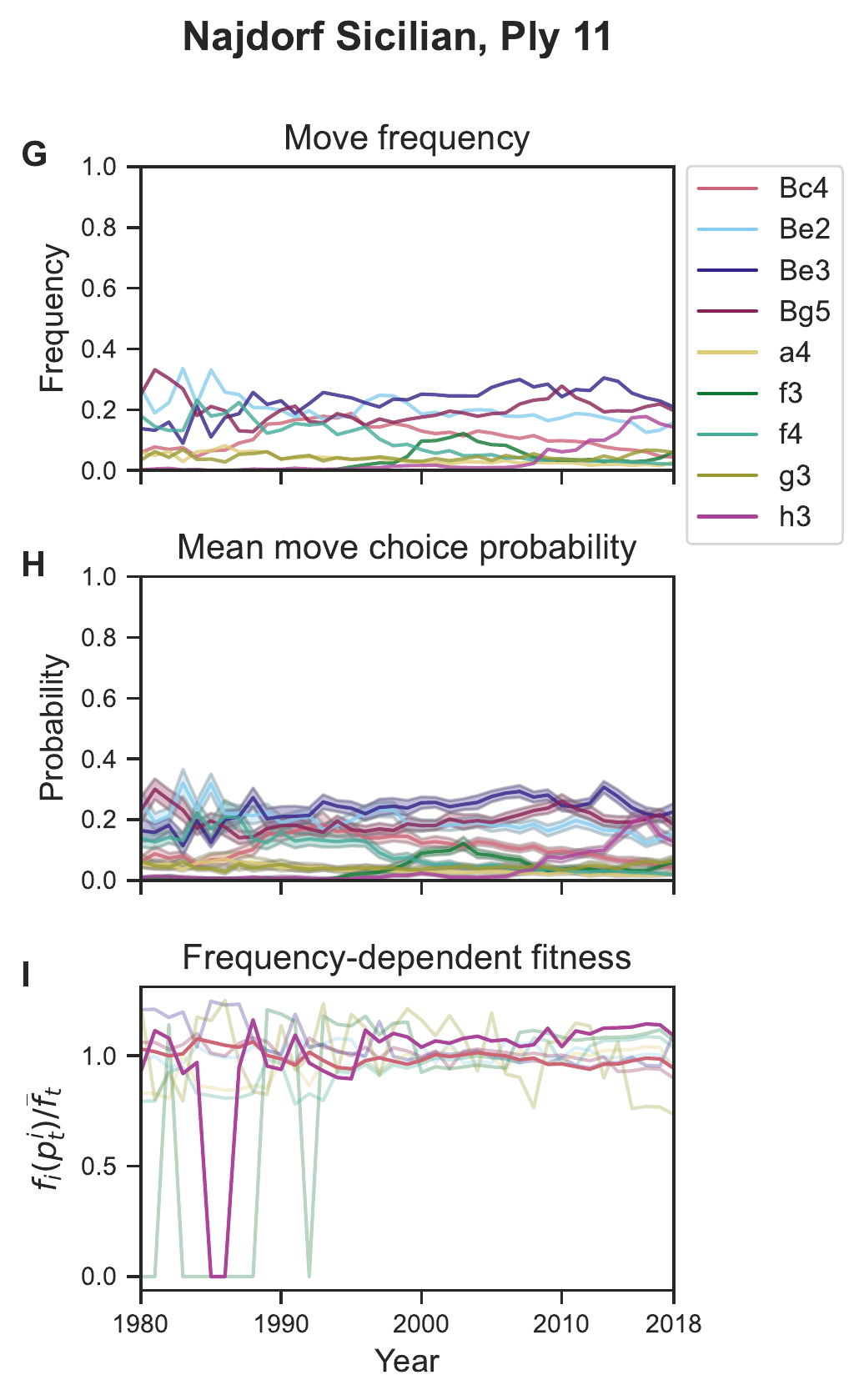}\label{subfig:model c}}    
    \caption{Dirichlet-multinomial model fits for move choice in three different positions: the Queen's Pawn opening at ply 2 (\textbf{1.\ d4}), the Caro-Kann opening at ply 5 (\textbf{1.\ e4 c6 2.\ d4 d5}), and the Najdorf Sicilian at ply 11 (\textbf{1.\ e4 c5 2.\ Nf3 d6 3.\ d4 cxd4 4.\ Nxd4 Nf6 5.\ Nc3 a6}). Panels A, D, and G show move frequencies $x_t^i/N_t$. Panels B, E, and H show posterior means of probabilities of move choice in the year $t$, with grey lines marking the range containing the middle 98\% of the posterior density. Panels C, F, and I show frequency-dependent fitness $f_i(x^i_t/N_t)/\bar f_t$ of moves over time, with the values computed using posterior medians of the $f_i$. (A) Move frequencies, Queen's Pawn, ply 2. (B) Mean move choice probability, Queen's Pawn, ply 2. (C) Frequency-dependent fitness, Queen's Pawn, ply 2. (D) Move frequencies, Caro-Kann, ply 5. (E) Mean move choice probability, Caro-Kann, ply 5. (F) Frequency-dependent fitness, Caro-Kann, ply 5. (G) Move frequencies, Najdorf Sicilian, ply 11. (H) Mean move choice probability, Najdorf Sicilian, ply 11. (I) Frequency-dependent fitness, Najdorf Sicilian, ply 11. The curves for the ``other'' category are omitted in all plots as the category is too rare to give meaningful results. The model was fitted for years 1980-2019, and the move fitnesses are estimated for all years except 2019.}
    \label{fig:model fitness dynamics}
\end{figure}

\begin{figure}[bht]\centering
    \includegraphics[width = 0.8\textwidth]{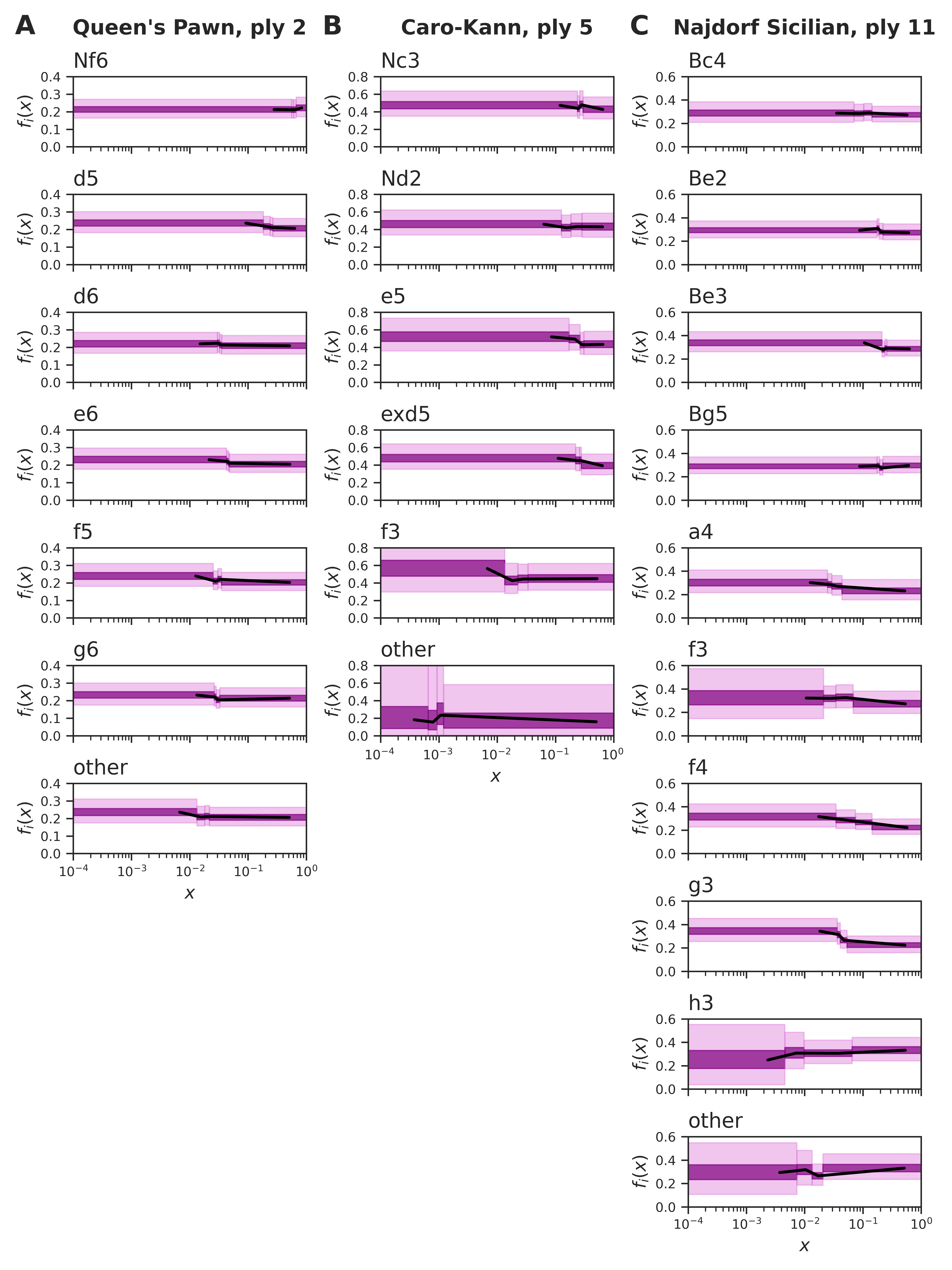} 
    \caption{Estimated frequency-dependent fitness functions $f_i$. The black line connects the posterior medians for the four constant segments, bright purple shows regions containing 60\% of the posterior density, and light purple shows regions containing 98\% of the posterior density. (A) Queen's Pawn, ply 2. (B) Caro-Kann, ply 5. (C) Najdorf Sicilian, ply 11.}
    \label{fig:fitness curves}     
\end{figure}

\begin{figure}[bht]\centering
    \includegraphics[width = 0.8\textwidth]{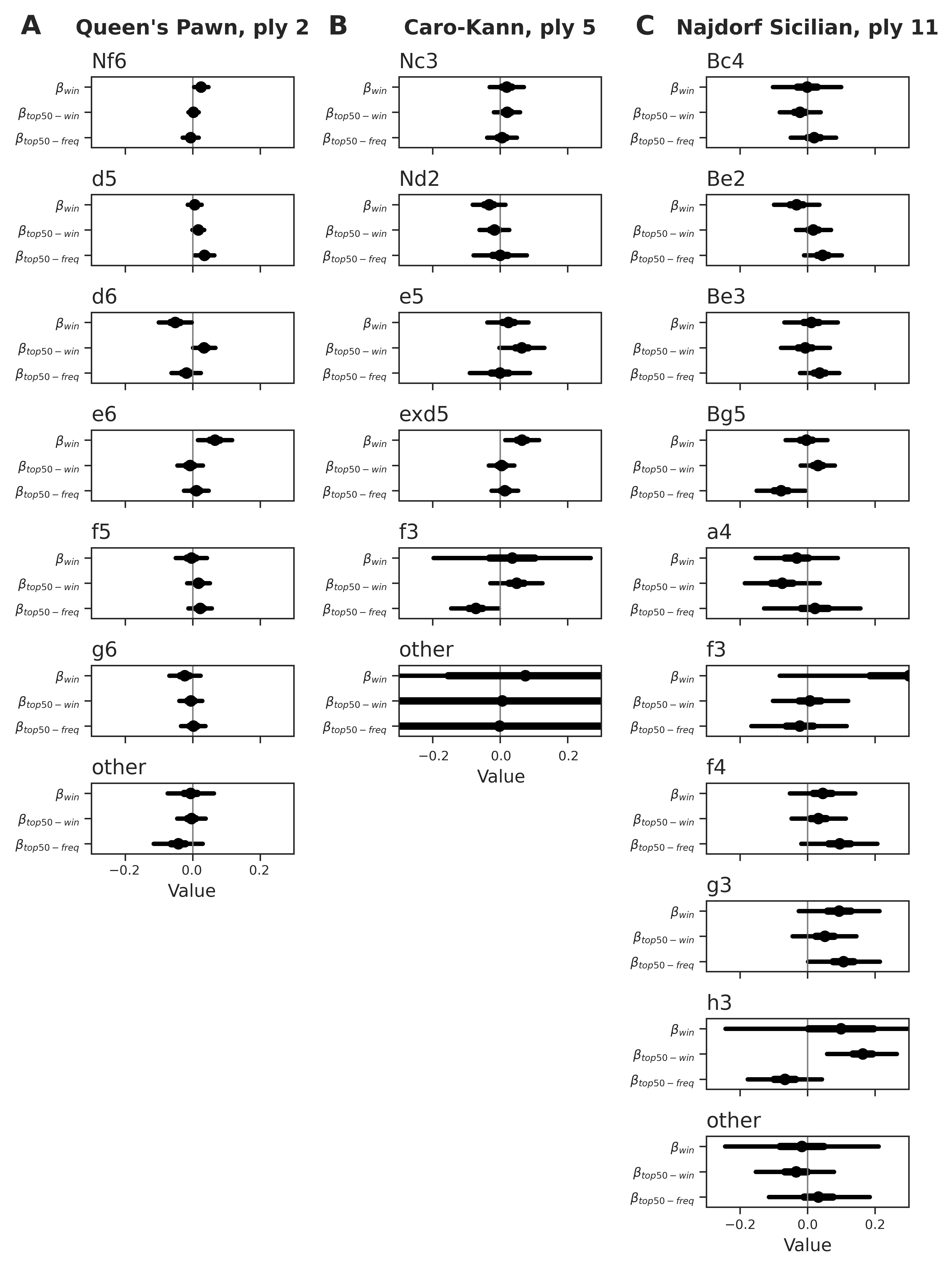}     
    \caption{Estimated coefficients $\boldsymbol \beta_i$. A point marks the posterior median, the thick line marks the region containing 60\% of the posterior density, and the thin line shows the region containing 98\% of the posterior density. 
    The coefficients presented are: $\beta_{\text{win}}$, the effect of the average outcome of games in the year previous to that in which a given move was played; $\beta_{\text{top50-win}}$, the effect of the average outcome of games involving players in the top 50 in the previous year; and $\beta_{\text{top50-freq}}$, the effect of the frequency of a given move in games involving players in the top 50 in the previous year (see \cref{subsec:model def}). (A) Queen's Pawn, ply 2. (B) Caro-Kann, ply 5. (C) Najdorf Sicilian, ply 11.}
    \label{fig:betas}     
\end{figure}

\begin{figure}[bth]
    \centering
    \includegraphics[width=0.9\linewidth]{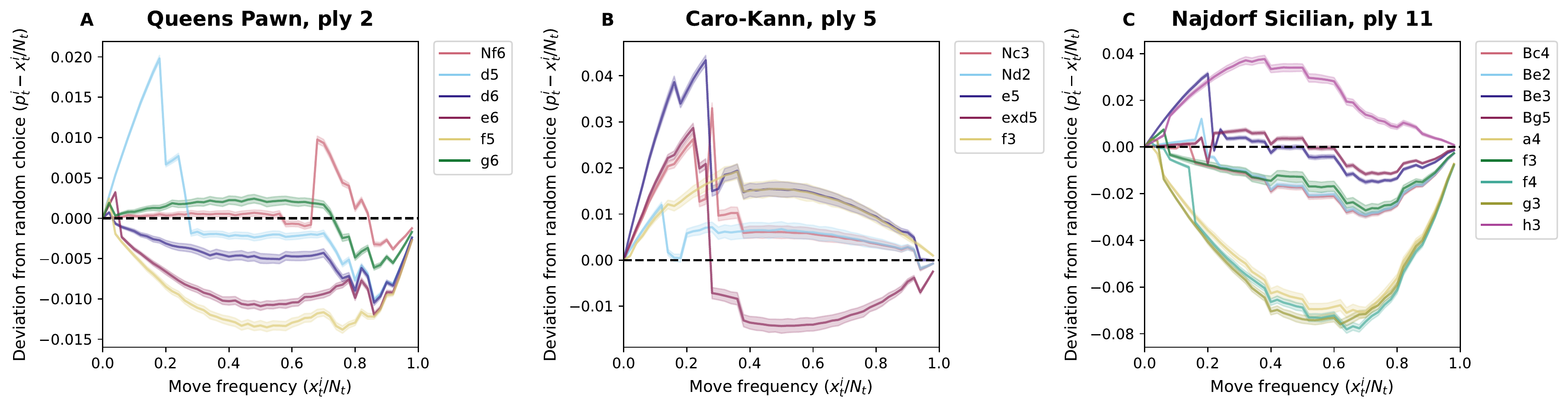}
    \caption{Dependence of move choice probability on strategy frequency. For each strategy, the corresponding curve shows the deviation from random choice of posterior move choice probability in year $t+1$ as the frequency of that strategy in year $t$ varies from 0 to 1. The values were computed using 1000 samples from the posterior for each initial move frequency in year $t$, as described in \cref{supp:counterfact}. Shading around the curves corresponds to 98\% bootstrap confidence intervals for the mean. The curves for the ``other'' category are omitted in all plots, as these are too rare to give meaningful results. (A) Queen's Pawn, ply 2. (B) Caro-Kann, ply 5. (C) Najdorf Sicilian, ply 11.}
    \label{fig:counterfactual curves}
\end{figure}

\Cref{fig:model fitness dynamics} shows the original frequency data, the move choice probabilities as estimated by the model, and estimates of frequency-dependent fitness ${f_i}'(x^i_t/N_t) = f_i(x^i_t/N_t)/\bar f_t$ of moves over time, as defined in \cref{eq:fbar,eq:fprime}. Comparing the first and second rows of panels in \cref{fig:model fitness dynamics}, our model fits the data well, with estimated move choice probabilities (panels B, E, H) matching the actual move frequencies (panels A, D, G). The estimates of the parameters $f_i$ and $\boldsymbol \beta_i$ are presented in \cref{fig:fitness curves,fig:betas}, respectively. For point estimates, the posterior median is used, and for quantifying uncertainty, we report posterior 1\% and 99\% quantiles for each estimate. In our analysis, we focus on effects $\boldsymbol \beta$ for which the middle 98\% of the distribution does not contain zero and on significant effects that have reasonable justifications in chess literature or history. Finally, \cref{fig:counterfactual curves} illustrates frequency dependence in the choice of strategies using posterior predictive sampling. We discuss \cref{fig:counterfactual curves} in detail below.

\subsection{Frequency dependence: Queen's Pawn opening}\label{subsec:results:fd}

Considering the responses to the Queen's Pawn opening in \cref{fig:model fitness dynamics}A, from 1980 to 2005, the move \textbf{d5} is, on average, increasing in popularity, with this trend reversing after 2005. The move \textbf{Nf6} shows the opposite dynamics. In fact, in World Championship matches of 2016, 2018, and 2021, players responded with \textbf{Nf6} in all but one game in this position \citep[see e.g.][]{2021wc}. Gradual changes can be caused by cultural drift \citep{bentley2007regular} or changes in metastrategy. However, our model suggests that transmission biases may play a role as well. In particular, the values of the fitness functions for \textbf{d5} and \textbf{Nf6} observed in \cref{fig:model fitness dynamics}C are higher when they are are at lower frequencies. The plots of frequency-dependent function functions $f_i(x)$ for $x$ from $0$ to $1$ are shown in \cref{fig:fitness curves}A, and there is a downward slope in the values of $f_i(x)$ characteristic of negative \emph{frequency-dependent bias}, or anti-conformity. Win rates or features related to top-50 players appear to have no effect on the choice of \textbf{d5} or \textbf{Nf6} (\cref{fig:betas}A). The other strategies are played in only a small proportion of games, and for those strategies, it may be hard to distinguish meaningful effects from statistical artifacts.

To further understand the nature of frequency dependence, we plot expected deviations of move choice probability $\mathbb E[p^i_{t+1}]$ from random choice ($\mathbb E[p^i_{t+1}] = p^i_t$) for initial frequencies $x^i_t/N_t = 0,0.02,\ldots,0.98, 1$, keeping other variables constant (see \cref{supp:counterfact} for a detailed description of the calculation). For the Queen's Pawn opening, this plot appears in \cref{fig:counterfactual curves}A. The choice of move \textbf{d5} clearly has negative frequency dependence, as it is chosen with probability higher than what is expected under random choice when its frequency is low and with lower probability when its frequency is high, with deviations from random choice as large as 1.9\%. Similar behavior can be seen for the move \textbf{Nf6}.

\subsection{Success bias: Caro-Kann}\label{subsec:result:success}

In the Caro-Kann opening, the move \textbf{exd5} is used less and less in more recent years (\cref{fig:model fitness dynamics}D). The plot of move fitnesses in \cref{fig:model fitness dynamics}F and the choice probability plot in \cref{fig:counterfactual curves}B suggest that negative frequency-dependent dynamics play a role in determining this behavior.
However, the functions $f_i$ are not the only determinants of move frequencies in our model; the coefficients $\boldsymbol\beta_i$ shown in \cref{fig:betas}B suggest that the choice to play the move \textbf{exd5} is affected by the win rate in the population, indicating \emph{success bias}. The decrease in the frequency of \textbf{exd5} then comes from many players losing after playing this move (see \cref{fig:input features}K). Indeed, computer engines have shown that the move \textbf{e5} provides the strongest winning probability for the player, while after \textbf{exd5} the opponent can ``equalize'' the position and take over the game \citep{carokann}.

\subsection{Prestige bias: Najdorf Sicilian}\label{subsec:result:prestige}

In the case of the Najdorf Sicilian, in \cref{subsec:streamplots}, we highlighted \textbf{h3} as a recent strong trend. The frequency-dependent fitness function $f_{\textbf{h3}}$ shows that there is no negative frequency-dependent bias for a choice of \textbf{h3} (\cref{fig:fitness curves}C); in fact, \cref{fig:counterfactual curves}C shows that \textbf{h3} is, on average, chosen with probability greater than random choice at every value of the frequency in the previous year. This result suggests that the move is a genuine innovation, becoming more popular ``on its own merit'' and not because of frequency-dependent trends. The coefficient for the win rate among the top 50 players, $\beta_{\text{\textbf{h3}},\text{top50-win}}$ is large (\cref{fig:betas}C), meaning that the increase in the frequency of \textbf{h3} could possibly be due to a trend started by elite players, which then led to wider adoption and development of theory. We conclude that the choice to play \textbf{h3} is subject to \emph{prestige bias}. In chess literature, side pawn pushes such as \textbf{h3}, \textbf{h4}, \textbf{a3}, and \textbf{a4} in various positions are ideas introduced by strong chess engines \citep[Ch.~9]{alphazero} in the most recent decade. This trend may explain why top players, who often have teams analyzing engine suggestions for them, have been adopting the move \textbf{h3}, subsequently influencing the general population.

\subsection{Game sample size \texorpdfstring{$N_s$}{Ns}}\label{subsec:nsamp}

Finally, we address the way our model characterizes the variance of move counts in the data. As we have discussed in \cref{subsubsec:model:fitness}, the mean fitness $\bar f_t$ controls the variance of $x_{t+1}^i$ conditional on model parameters and $x_t^i$. Mathematically, this influence can be seen as follows. As a shorthand, let

\begin{equation} 
  p_i = \frac{f_i(x_t^i/N_t)\, x_t^i}{\sum_{j=1}^K f_i(x_t^j/N_t)\, x_t^j}
\end{equation}
be the ``frequencies'' of strategies assuming no effect of prestige or success biases. Then the variance of $x_{t+1}^i$ is \citep[][p.~81]{johnson1997discrete}:
\begin{equation}\label{eq:ns variance}
  \Var_{\text{DM}}(x_{t+1}^i \mid x_t^i) = N_{t+1} p_i (1-p_i) \left(\frac{N_{t+1} + N_t \bar f_t}{1 + N_t \bar f_t}\right).
\end{equation}
The last term of \cref{eq:ns variance} is a decreasing rational function of $\bar f_t$, so $\Var_{\text{DM}}(x_{t+1}^i \mid x_t^i)$ decreases as $\bar f_t$ grows.

In the fitted models, the mean fitness $\bar f_t$ is consistently below 1 for all three positions considered, equal to $\sim 0.22$ for the Queen's Pawn, ply 2 position (approximately constant over time), $\sim 0.3$ for Caro-Kann, ply 5, and $\sim 0.45$ for Najdorf Sicilian, ply 11 (\cref{fig:fbar}A). That we have observed $\bar f_t < 1$ can be interpreted in relation to players' behavior. Mechanistically, our model describes players observing move counts in a previous year, adjusting their preferences because of transmission biases, and then selecting a move with \emph{higher variance} than what is expected if $\bar f_t=1$, corresponding to multinomial choice. 
We define \emph{game sample size} $N_s(t) = \bar f_t N_t$ to be the number of games in the population at time $t$ that achieves the same value for the variance $\Var_{\text{DM}}(x_{t+1}^i \mid x_t^i)$ as in \cref{eq:ns variance} under the condition $\bar f_t = 1$. Indeed, with game sample size defined as $N_s(t) = \bar f_t N_t$, \cref{eq:ns variance} becomes

\begin{equation}
  \Var_{\text{DM}}(x_{t+1}^i \mid x_t^i) = N_{t+1} p_i (1-p_i) \left(\frac{N_{t+1} + N_s(t)}{1 + N_s(t)}\right),
\end{equation}
so that now a mechanistic interpretation of our model consists of players observing move counts in a population of size $N_s(t)$, adjusting their preferences according to transmission biases, and then choosing the strategy according to a multinomial distribution.

As the game progresses from early to later positions, the players sample a higher fraction of all games in their decision-making process (\cref{fig:fbar}). Possibly, the fraction of games sampled by players is low for early positions because tens of thousands of professional games each year start with a move \textbf{d4} (\cref{fig:input features}A), and it is likely that players cannot monitor all of these games. However, a player who specializes in playing the Najdorf Sicilian may pay attention to a larger proportion of games involving this opening, because the total number of games to analyze is much smaller for ply 11 in the Najdorf Sicilian (\cref{fig:input features}C) than in the Queen's Pawn at ply 2 (\cref{fig:input features}A).

\section{Discussion}\label{sec:discussion}

We have developed a population-level model for the influence of transmission biases on move choice in chess. We have shown that many of the moves analyzed are under negative frequency-dependent cultural selection, having higher fitness and being favorably selected with probability greater than random choice at lower frequencies (\cref{fig:model fitness dynamics,fig:counterfactual curves}). This result suggests that anti-conformity is important in the transmission of chess opening strategies. In addition, our model is able to identify moves for which other factors play a role: the dynamics of \textbf{h3} in the Najdorf Sicilian are affected by the win rate among the top 50 players (\cref{fig:betas}C), indicating the presence of prestige bias, and the choice of \textbf{exd4} in the Caro-Kann suggests success bias (\cref{fig:betas}B). 

We have also inferred absence of significant success bias for many strategies, consistent with our discussion in \cref{sec:culture}: a win in chess is conditional on strong performance at every move, so making decisions about the opening based on the average eventual outcome may not be the best choice from many positions. Similarly, following choices of top players would be effective only if a strong continuation were found. Support for our findings of strong success bias in the Caro-Kann and prestige bias in the Najdorf Sicilian comes from information commonly known to professional chess players, such as new insights from extensive computer analysis, or new styles of play introduced by computer players.

In addition to measuring transmission biases, we have introduced a concept of ``game sample size'' $N_s$ that appears naturally from the analysis of game counts (\cref{subsec:nsamp}). $N_s$ can be interpreted as the number of games that players observe when making use of social information. We have shown that later positions have a greater ratio $N_s/N$, which could mean that players use more complete information when positions become more complicated and less standardized (\cref{fig:fbar}).

The estimated game sample size relates to several theoretical concepts. First, from the perspective of population genetics, $N_s$ is equivalent to variance effective population size $N_e(t) = \bar f_t N_t$ used to account for overdispersed allele counts relative to a standard Wright-Fisher model \citep[Ch.~3]{ewens,caballero2020quantitative}. Second, theoretical studies of conformity typically involve individuals sampling role models from the population and making a choice based on this sample \citep{boyd_richerson, denton2020cultural,fpm_conformity}. The number of role models is usually taken to be equal to a small number such as 3, which is much smaller than the population size. The value of $N_s$ can be seen as relating to these theoretical models, measuring how many role models are sampled from the population.

\subsection{Related and complementary work}

Our model complements other recent work on measuring the strength of transmission biases in cultural datasets of competitive activities, such as the studies by \cite{beheim2014strategic} on Go, \cite{miu2018innovation} on programming contests, and \cite{mesoudi2020cultural} on football strategy. \cite{beheim2014strategic} employed multilevel logistic regression to study social and individual learning involved in the board game Go. They observed strong success bias and \emph{positive} frequency-dependence for the choice of one of the opening moves. Positive frequency-dependence in Go and negative frequency-dependence in chess could be connected to the differences in the communities around each game. Among board games, chess is unique in its use of computer engines. Computer chess engines became widely available to elite players starting from the late 1990s, revolutionizing tournament preparation. Finding the best response in a position or solving a chess puzzle became possible in a matter of seconds rather than hours or days. Post-game analysis now helps players quickly identify and address their weaknesses, which means that players can no longer ``catch'' many opponents with the same ``trick.'' Playing into popular lines can also lead to positions in which the opponent has the most preparation. In contrast, the space of possible moves in the opening is much larger in Go, and computers have reached human level only in the most recent decade. Hence, the effectiveness of studying a \emph{particular} position in Go is diminished, and players may choose to conform to a popular strategy for their first move and hope to outplay the opponent later in the game.

Transmission biases and social learning strategies in various games have also been measured in field observations and experiments \citep[e.g.\ ][]{aplin2017conformity, barrett2017pay, deffner2020dynamic, vale2017acquisition, canteloup2021processing}. 
Studies using experimental data typically involve models that estimate parameters for each observed individual or category of individuals, whereas we focus on analysis of large-scale population-level data. Still, some aspects of such models are similar to our Dirichlet-multinomial approach. For example, in the experience-weighted attraction (EWA) model employed by \cite{barrett2017pay} to analyze social learning in Capuchin monkeys (also used in \cite{canteloup2021processing} and \cite{deffner2020dynamic}), decisions are influenced by a convex combination of functions representing individual and social learning, and different social biases are encoded in a multiplicative way similar to \cref{eq:model alpha} in our model. This similarity suggests that our model could potentially be modified to model move choice of each player via a Dirichlet-multinomial likelihood, enabling comparisons of learning modalities between individual players.

Frequency-dependent selection has previously been measured by \cite{plotkin2022} in other large datasets such as baby name statistics and dog breed popularity data. These authors focused on modeling ``exchangeable'' entities, for which selection acts on every variant in exactly the same way. They estimate a single fitness function that is shared by every cultural variant and that characterizes average frequency-dependence in the population. Chess differs from such contexts in that it contains the concept of a ``win.'' Each chess move leads to a different position, altering the winning chances of each player, so that strategies at different stages of the game are dependent. Thus, our model assigns a separate fitness function to each individual strategy, treating strategies as \emph{nonexchangeable}.

Our model also extends the multinomial model of \cite{plotkin2022} by incorporating the Dirichlet distribution into the model likelihood. This approach has a clear mechanistic interpretation in terms of players' behaviors and allows us to perform efficient Bayesian inference of model parameters.
Statistical models of count data based on the Dirichlet-multinomial likelihood are known in many related areas, including linguistics \citep[e.g.][]{madsen2005modeling}, human genetics \citep{wang2023human}, molecular ecology \citep{harrison2020dirichlet}, and microbiome data analysis \citep[e.g.][]{osborne2022latent,wadsworth2017integrative}.

\subsection{Caveats}

The parameters of our model can be represented in two different ways. One uses $k$ fitness functions $f_i$ that are only constrained to be non-negative and are naturally suited to Bayesian inference. Another uses $k$ functions $f_i'$ (\cref{eq:fprime}) that are required to sum to one, together with the mean fitness $\bar f_t$ (\cref{eq:fbar t}). While estimates of $f_i$ (\cref{fig:fitness curves}) show presence of frequency-dependent dynamics, it is hard to characterize strength and significance of frequency-dependence using the values of $f_i$. To understand strength and significance of frequency-dependent effects we plot the growth rates $f_i'$ of strategies (\cref{fig:model fitness dynamics}) and compute expected deviation of move counts from random choice (\cref{fig:counterfactual curves}). Other analyses could potentially be used, for example evaluating whether the function $f_i$ is significantly different from a constant function.

Another statistical issue that could affect our inferences is the possibility of correlated input features, so that the $\boldsymbol\beta$ coefficients might not be easily identifiable. However, features for the games played by the top-50 players show behaviors that differ from those of the total population of around 15,000 players (\cref{fig:input features}). Thus, for the factors we consider, it appears that distinguishing the influence of the top-50 players from a general influence of master-level players is indeed possible.

Our model incorporates only a subset of possible features that could be relevant to move choice, such as highly developed theory or objective strength as determined by computer evaluation (\cref{sec:culture}). However, the presence of significant success bias and prestige bias could correspond to mechanisms of social learning \emph{about} these other features. For example, suppose a player observes several successful games in top tournaments with \textbf{h3} in the Najdorf Sicilian, and then studies the move. The player could learn about the enthusiasm of modern computer engines for this move and could incorporate it into future play. For our model, this mechanism is indistinguishable from the player simply copying a successful move. This reasoning about a player's mechanistic evaluation process suggests a potential direction for further modeling that would incorporate varying individual behaviors and knowledge about position evaluation.

\subsection{Conclusion}

Data from the last five decades of high-level chess games can be evaluated in the context of cultural transmission and evolution. We have shown that the cultural ``features'' of transmission can be measured from move choice decisions in various positions by professional players. In particular, we have inferred influences of frequency-dependent bias, success bias (win rate), and prestige bias (the use of the move by the very top players). The prevalence of anti-conformity and the lack of strong success bias for many strategies reflects the nature of opening play in chess, which involves extensive preparation and assessment of opponents' likely preparation. We have also connected the presence or absence of transmission biases with chess theory. The fact that many of our quantitative results correspond to ideas well-known to professional chess players suggests that our modeling could be useful to chess analysts and historians. In particular, many qualitative explanations are available for the popularity of certain strategies, and a quantitative evaluation of move frequency dynamics could help test the narratives familiar to chess players with statistical evidence. More broadly, our statistical approach could potentially be used to complement the historical study of cultural trends in other games that contain discrete choices, or even in other cultural domains in which circumscribed discrete data are recorded.

\vskip .3cm
{\small
\noindent \textbf{Data and code.} The code to generate figures in this paper and links to access the dataset are available at \href{https://github.com/EgorLappo/cultural_transmission_in_chess}{\texttt{github.com/EgorLappo/cultural\_transmission\_in\_chess}}.

\noindent \textbf{Funding.} We acknowledge NSF grant BCS-2116322 and grant 61809 from the J.\ T.\ Templeton Foundation for support.

\noindent \textbf{Acknowledgments.}  We thank Sharon Du for suggesting the problem and Kaleda Denton, Bret Beheim, and an anonymous reviewer for helpful comments on the manuscript.
}

\printbibliography

\clearpage

\noindent\textbf{\LARGE Supplementary information}

\appendix

\numberwithin{equation}{section}
\setcounter{figure}{0}
\renewcommand\thefigure{S\arabic{figure}}
\renewcommand\thesection{S\arabic{section}}
\renewcommand\thetable{S\arabic{table}}

\section{Model fitting}\label{supp:stan info}

In this section we give details of the fitting procedure and convergence diagnostics for the model of move choice defined in \cref{subsec:model}. For each of the three strategies discussed in \cref{sec:model analysis}, the model was fitted using the \texttt{cmdstanr} R package, running 4 parallel chains for 10,000 sampling iterations with 2,000 iterations for warm-up. There were no divergent transitions, and reported E-BFMI values were above 0.85 for all chains, showing that the models explored the posterior well \citep{betancourt2017conceptual, mcmc_diagnostics}. 

\begin{figure}[bth]
    \centering
    \includegraphics[width=0.7\linewidth]{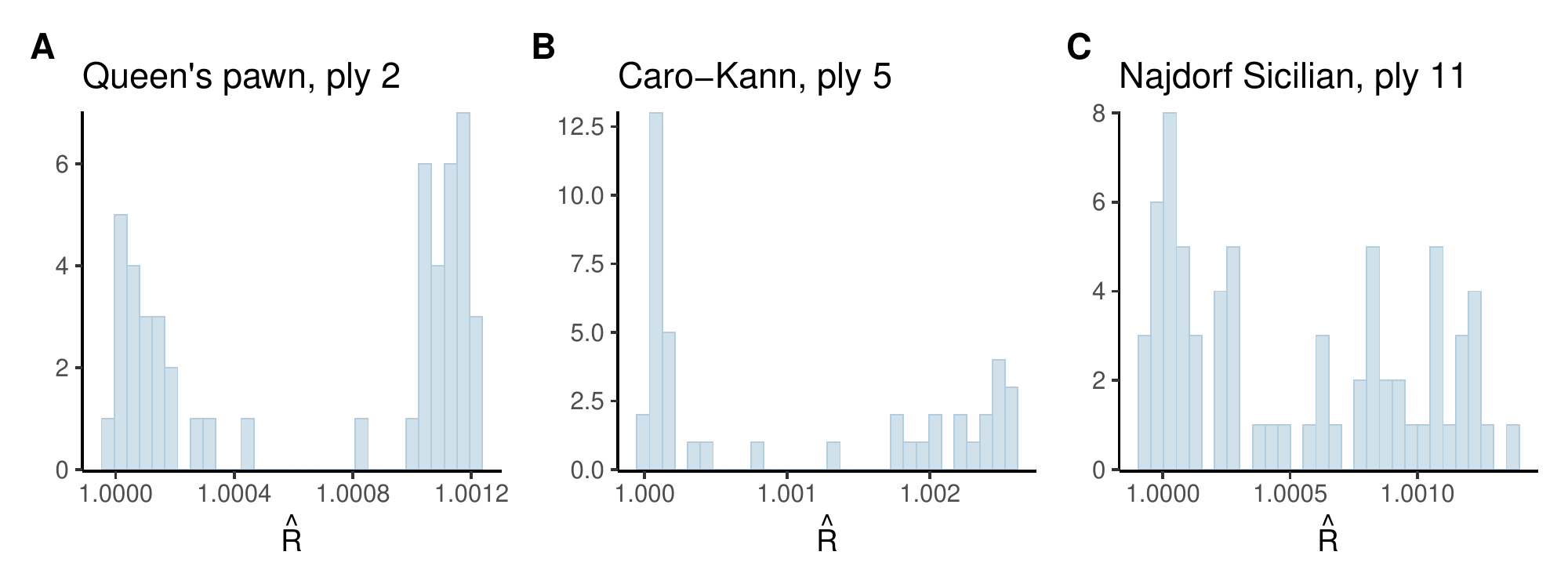}
    \caption{Estimates of $\hat R$ convergence diagnostics for Hamiltonian Monte Carlo model fits. (A) Histogram of the $\hat R$ values for the $7\times 7 = 49$ parameters of the Queen's Pawn opening model. (B) Histogram of the $\hat R$ values for the $6\times 7 = 42$ parameters of the Caro-Kann opening model. (C) Histogram of the $\hat R$ values for the $10 \times 7 = 70$ parameters of the Najdorf Sicilian opening model.}
    \label{fig:rhat}
\end{figure}

Convergence of the chains was confirmed by visual inspection of traceplots and rank histograms, as well as using the $\hat R$ diagnostic \citep{gelman1992inference, vehtari2021rank}. The histograms of $\hat R$ values for all $7k$ model parameters (where $k$ is the number of strategies) are shown in \cref{fig:rhat}, and all of them are below 1.01, which is considered sufficient evidence for convergence \citep{mcmc_diagnostics}.

\clearpage

\section{Model input data}\label{supp:input features}

\begin{figure}[bth]
    \centering
    \includegraphics[width=0.9\linewidth]{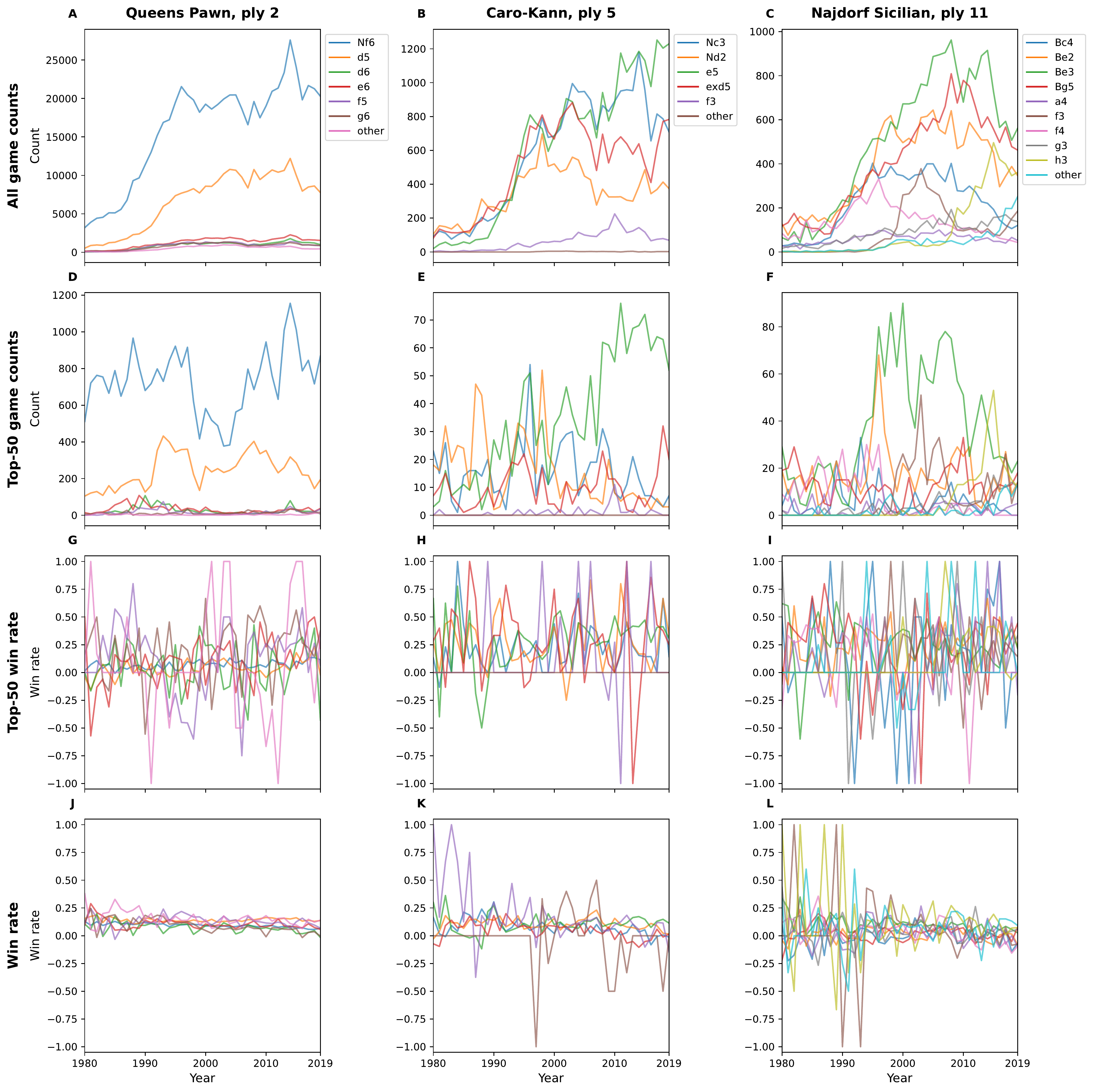}
    \caption{Input data for the Dirichlet-multinomial model for the three positions discussed in \cref{sec:model analysis}. (A, B, C) Game counts by strategy for the Queen's Pawn, Caro-Kann, and Najdorf Sicilian positions respectively. (D, E, F) Counts of games played by the top-50 players for the Queen's Pawn, Caro-Kann, and Najdorf Sicilian positions, respectively. (G, H, I) Win rates for the Queen's Pawn, Caro-Kann, and Najdorf Sicilian positions, respectively. (J, K, L) Win rates in games played by the top-50 players for the Queen's Pawn, Caro-Kann, and Najdorf Sicilian positions, respectively.}
    \label{fig:input features}
\end{figure}

\Cref{fig:input features} shows the data that was input into the model for each of the three strategies discussed in \cref{sec:model analysis}: the raw strategy counts $x_t^i$, strategy counts among the top-50 players, strategy win rates in the total population, and win rates among the top-50 players. 

\clearpage

\section{Frequency-dependence characterization}\label{supp:counterfact}

In this section, we describe in detail the procedure used to generate counterfactual predictions for move choice probability in \cref{fig:counterfactual curves}. We sample move choice probabilities for year $t+1$ from the posterior for our model while keeping as many variables as possible constant in year $t$. We fix the number of games at $N_t = 100,000$ and additionally set the linear features (win rate, win rate among top-50 players, and game count among top-50 players) to be constant and equal to their time averages,

\begin{equation*}
    \widetilde{\boldsymbol y}^i = \frac{1}{40} \sum_{t=0}^{39} \boldsymbol y^i_t.
\end{equation*}
As these linear features were standardized to to have zero mean before being input into the model, we must have $\widetilde{\boldsymbol y}^i=0$ and therefore $\exp(\boldsymbol \beta_i \cdot \widetilde{\boldsymbol y}^i) = \exp(0) = 1$ in \cref{eq:model alpha mod}. 

Next, for each strategy $i$, we generate a sequence of counts $x^i_t$ from $0$ to $N_t$. We set the counts of all other strategies to $x^j_t = (N_t - x^i_t)/(k-1)$, where $i\neq j$ and $k$ is the number of strategies, which allows for fractional values. This choice of counts keeps the other strategy counts equal while varying the frequency of a strategy of interest. 

Finally, we use these input data to sample model predictions of move choice probabilities $p_t^i$ from the fitted posterior distribution. We first sample coefficients $c^i_j$ from the posterior, compute coefficients $\boldsymbol\alpha$ using \cref{eq:model alpha mod}, and then sample move probabilities $p_t$ from $\text{Dirichlet}(\boldsymbol\alpha)$. We repeat this procedure 1,000 times. As a result, for each strategy $i$ and each initial frequency $x_t^i$, we obtain 1,000 samples of move choice probability $p_t^i$ as estimated by our model. Recall that when strategies are chosen randomly among games in the previous year, the probability of choosing strategy $i$ is equal to the frequency of $i$ in the population, $x_t^i/N_t$. Comparing our estimates of $p_t^i$ to $x_t^i/N_t$, the expectation under random choice, allows us to characterize frequency-dependent effects in our model. 

\clearpage

\section{Plots of mean fitness \texorpdfstring{$\bar f_t$}{fbar} and game sample size \texorpdfstring{$N_s$}{Ns}}\label{supp:fbar}

\begin{figure}[bth]
    \centering
    \includegraphics[width=0.9\linewidth]{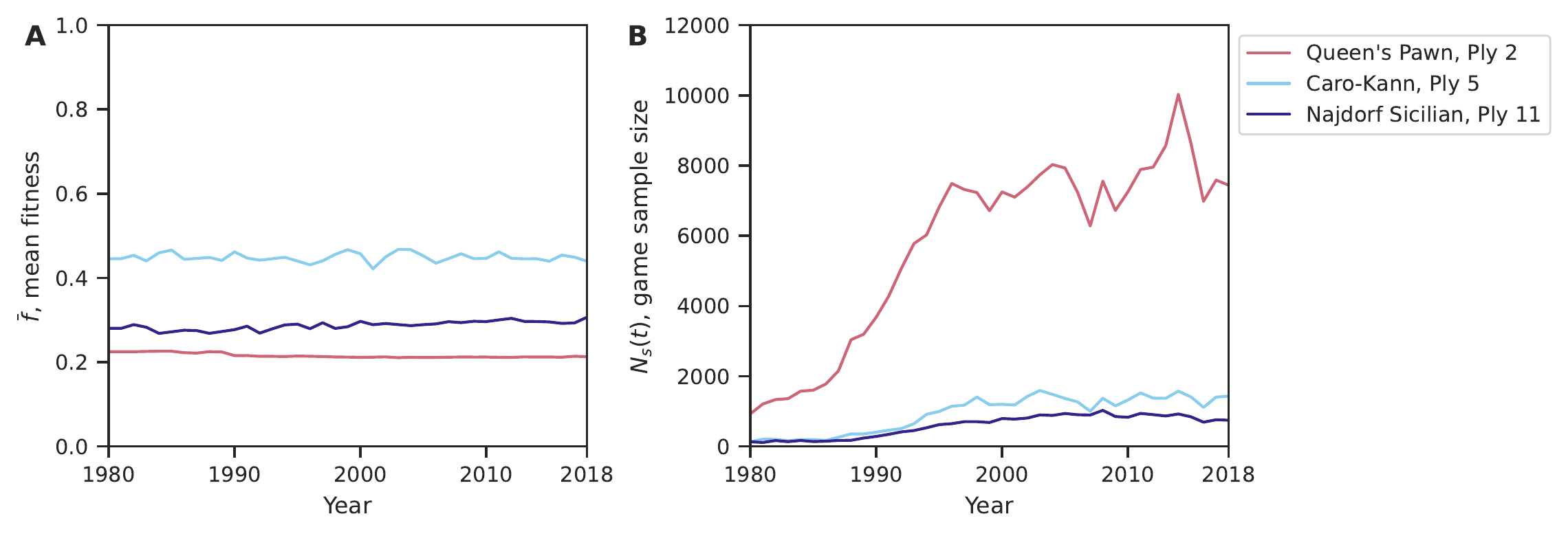}
    \caption{Mean fitness and game sample size estimated by our model for three positions: Queen's Pawn opening, Caro-Kann opening, and Najdorf Sicilian. (A) Mean fitness $\bar f_t$. (B) Game sample size $N_s(t) = N_t \bar f_t$.}
    \label{fig:fbar}
\end{figure}

Plots of mean fitness $\bar f_t$ and game sample size $N_s(t) = N_t \bar f_t$ (\cref{subsec:nsamp}) estimated from the Dirichlet-multinomial model are shown in \cref{fig:fbar}.

\end{document}